\documentclass{jpp}
\usepackage{graphicx}

\usepackage[utf8]{inputenc}
\usepackage[T1]{fontenc}
\usepackage{amsmath}
\usepackage{amssymb}
\usepackage{bm}
\usepackage{url}
\usepackage{color}

\newcommand{\EQ}{\begin{equation}}
\newcommand{\EN}{\end{equation}}
\newcommand{\EQA}{\begin{eqnarray}}
\newcommand{\ENA}{\end{eqnarray}}

\newcommand{\Eq}[1]{Eq.~(\ref{#1})}

\newcommand{\Fig}[1]{Figure~\ref{#1}}

\newcommand{\Tab}[1]{Table~\ref{#1}}

\newcommand{\bra}[1]{\langle #1\rangle}

{}
{}

{}
{}

{}
{}
{}
{}
{}
{}
{}
{}
{}
{}
{}
{}
{}
{}

{}

{}

{}
{}
{}

\newcommand{\xx}{\bm{x}}

\newcommand{\rr}{\bm{r}}

\newcommand{\BB}{\bm{B}}

\newcommand{\JJ}{\bm{J}}

\newcommand{\AAA}{\bm{A}}

\newcommand{\uu}{\bm{u}}

\newcommand{\nab}{{\bm{\nabla}}}

\newcommand{\dd}{{\rm d} {}}

\newcommand{\const}{{\rm const}  {}}

\def\Lu{\mbox{\rm Lu}}

\def\EEEE{{\cal E}}

\def\Brms{B_{\rm rms}}

\newcommand{\T}{\,{\rm T}}

\newcommand{\s}{\,{\rm s}}

\newcommand{\m}{\,{\rm m}}

\newcommand{\kg}{\,{\rm kg}}

\newcommand{\A}{\,{\rm A}}

\usepackage[normalem]{ulem}
\usepackage{xcolor}

\shorttitle{Hosking integral in Hall cascade}
\shortauthor{A. Brandenburg}

\title{Hosking integral in nonhelical Hall cascade}

\author{
Axel Brandenburg
}

\affiliation{
Isaac Newton Institute for Mathematical Sciences,
20 Clarkson Road, Cambridge CB3 0EH, UK; and
Nordita, KTH Royal Institute of Technology and Stockholm University,
Hannes Alfv\'ens v\"ag 12, SE-10691 Stockholm, Sweden
}

\begin{document}

\maketitle

\begin{abstract}
The Hosking integral, which characterizes magnetic helicity fluctuations
in subvolumes, is known to govern the decay of magnetically dominated
turbulence.
Here we show that, when the evolution of the magnetic field is controlled
by the motion of electrons only, as in neutron star crusts, the decay of
the magnetic field is still controlled by the Hosking integral, but now it
has effectively different dimensions than in ordinary magnetohydrodynamic
(MHD) turbulence.
This causes the correlation length to increase with time $t$ like
$t^{4/13}$ instead of $t^{4/9}$ in MHD.
The magnetic energy density decreases like $t^{-10/13}$, which is slower
than in MHD, where it decays like $t^{-10/9}$.
These new analytic results agree with earlier numerical simulations for
the nonhelical Hall cascade.
\end{abstract}

\section{Introduction}

The x-ray emission from neutron stars during the first hundreds
of years is believed to be powered by magnetic dissipation within their
outer crusts \citep{Gourgouliatos+18b}.
Since the ions are immobile in neutron star crusts, electric currents are
transported by electrons alone \citep{CL09}.
Their velocity is $\uu=-\JJ/en_e$, where $\JJ=\nab\times\BB/\mu_0$ is the
current density, $\BB$ is the magnetic field, $e$ is the elementary charge,
$n_e$ is the electron density, and $\mu_0$ is the permeability.
The evolution of $\BB$ is then governed by the induction equation where
the electromotive force $\uu\times\BB$ is given by $-\JJ\times\BB/en_e$.
The induction equation therefore takes the form \citep{GR92}
\begin{equation}
\frac{\partial\BB}{\partial t}=\nab\times\left(
-\frac{1}{en_e}\JJ\times\BB-\eta\mu_0\JJ\right),
\label{Hall}
\end{equation}
where $\eta$ is the magnetic diffusivity.
The $\JJ\times\BB$ nonlinearity in this equation leads to a cascade
toward smaller scales---similar to the turbulent cascade in hydrodynamics
turbulence \citep{GR92}.
This model is therefore referred to what is called the Hall cascade.
There has been extensive work trying to quantify the amount of dissipation
that occurs \citep{Gourgouliatos+16, Gourgouliatos+20, Gourgouliatos+18,
Igoshev+21, Anzuini+22}.
Idealized simulations in Cartesian geometry resulted in power law scaling
for the resistive Joule dissipation \citep[][hereafter B20]{B20}.
It depends on the typical length scale of the turbulence, the electron
density, the magnetic field strength, and the presence or absence of
magnetic helicity.
Denoting volume averages by angle brackets, the decay of the magnetic
energy density $\EEEE=\bra{\BB^2}/2\mu_0$ with time $t$ tends to
follow power law behavior, $\EEEE\propto t^{-p}$, where the exponent $p$
is smaller than in magnetohydrodynamic (MHD) turbulence.
Here, $\bra{...}$ denotes volume averaging over the spatial
coordinates $\xx$.
In the helical case, it was found that $p=2/5$, while for the nonhelical
case, B20 reported $p\approx0.9$.
The correlation length of the turbulence, $\xi$, increases with time
like $\xi\propto t^q$, where $q=2/5$ in the helical case, i.e., $q=p$,
and $q\approx0.3$ in the nonhelical case.
In the helical case, the exponent $2/5$ was possible to explain on
dimensional grounds by noting that the magnetic field does not correspond
to a speed (the Alfv\'en speed, as in MHD) with dimensions $\m\s^{-1}$
in SI units, but to a diffusivity with dimensions $\m^2\s^{-1}$.

The decay properties of the nonhelical Hall cascade were not yet
theoretically understood at the time.
In the last one to two years, however, significant progress has been
made in describing the decay of magnetically dominated turbulence, where
a new conserved quantity has been identified, which is now called the
Hosking integral \citep[see][for details regarding the naming]{ZSB22}.
The purpose of the present paper is to propose the scaling of the Hall
cascade under the assumption that it is governed by the constancy of the
Hosking integral, which now has different dimensions than in MHD.

\section{Hosking integral and scaling for the Hall cascade}

The Hosking integral $I_{\rm H}$ is defined as the {\em asymptotic limit of
the magnetic helicity density correlation integral ${\cal I}_{\rm H}(R)$
for scales $R$ large compared to the correlation length of the turbulence,
$\xi$}, but small compared to the system size $L$.
The original work on this integral is that by \cite{HS21}, who
subsequently applied it to the magnetic field decay in the early universe
\citep{HS22}; see also \cite{BKT15} and \cite{BK17} for earlier work
were inverse cascading of magnetically dominated nonhelical turbulence
was first reported.
The function ${\cal I}_{\rm H}(R)$ is given by
\begin{equation}
{\cal I}_{\rm H}(R)=\int_{V_R}\dd^3r \, \bra{h(\xx)h(\xx+\rr)},
\end{equation}
where $V_R$ is the volume of a ball of radius $R$ and $h=\AAA\cdot\BB$
is the magnetic helicity density with $\AAA$ being the magnetic vector
potential, so $\BB=\nab\times\AAA$.
We recall that $\bra{...}$ denotes averaging over $\xx$.
The function ${\cal I}_{\rm H}(R)$ is thus the integral over the volume
$V_R=4\pi R^3/3$.
For small $R$, it increases proportional to $R^3$, but for large $R$,
it levels off.
This is the value of $R$, which determines the Hosking integral
$I_{\rm H}\equiv{\cal I}_{\rm H}(R)$.
In practice, it is chosen empirically and must still be small compared
with the size of the domain; see \cite{HS21} for various examples and
\cite{ZSB22} for a comparison of different computational techniques for
obtaining ${\cal I}_{\rm H}(R)$.

What matters for the Hall cascade is the fact that the dimensions of $h$
are $[h]=[B]^2[x]$, and therefore the dimensions of ${\cal I}_{\rm H}$
and $I_{\rm H}$ are
\begin{equation}
[I_{\rm H}]=[B]^4[x]^5.
\end{equation}
However, as already noted in B20, using
$e=1.6\times10^{-19}\A\s$, $\mu_0=4\pi\times10^{-7}\T\m\A^{-1}$, and
$n_e\approx2.5\times10^{40}\m^{-3}$ for neutron star crusts, we have
$en_e\mu_0\approx5\times10^{15}\T\s\m^{-2}$, and therefore
\begin{equation}
\frac{B}{en_e\mu_0}=\frac{B}{5\times10^{15}\T}\,\frac{\m^2}{\s},
\end{equation}
which is why we say $B$ has dimensions of $\m^2\s^{-1}$.\footnote{
In MHD, by comparison, the ion density $\rho$ is a relevant quantity.
Using $\rho=10^{3}\kg\m^{-3}$ for solar surface plasmas,
and the identity $1\T=1\kg\s^{-2}\A^{-1}$, we have
$\mu_0=4\pi\times10^{-7}\T^2\s^2\m\kg^{-1}$, and therefore
$\rho\mu_0\approx3.5\times10^{-2}\T\s\m^{-1}$, or
$B/\sqrt{\rho_0\mu_0}=(B/3.5\times10^{-2}\T)\,\m\s^{-1}$,
which is why we say that in MHD, $B$ has dimensions of $\m\s^{-1}$.}
Therefore, the dimensions of $I_{\rm H}$ are
\begin{equation}
[I_{\rm H}]=[x]^{13}[\s]^{-4}.
\end{equation}
Thus, given that $I_{\rm H}=\const$ in the limit of small magnetic
diffusivity, a self-similar evolution must imply that all relevant length
scales, and in particular the magnetic correlation length $\xi(t)$,
must increase with time like $\xi\sim t^{4/13}$.
Since $4/13\approx0.31$, this is indeed close to the behavior $\xi\sim
t^q$ with $q\approx0.3$ found in Sec.~3.2 of B20 (their Run~B), as
already highlighted in the introduction of the present paper.

\begin{figure}
\centering
\includegraphics[width=0.9\columnwidth]{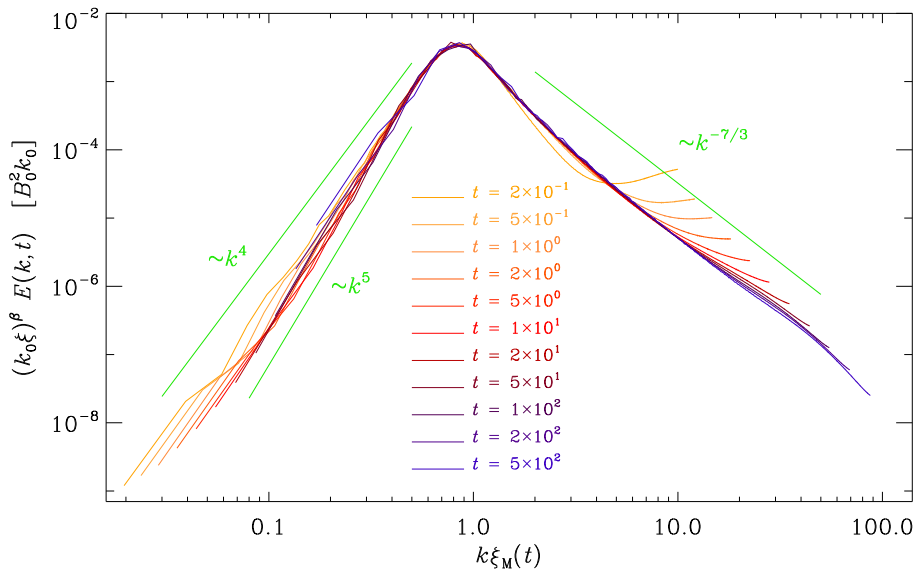}
\caption{
Compensated spectra for Run~B of B20, which corresponds to Run~B1
in the present paper.
Here, $\beta=1.7$ has been used as the best empirical fit parameter.
}\label{pkt_few_comp_Hf0_t2em5_k180a}
\end{figure}

To demonstrate that the energy spectra $E(k,t$ at different times
are indeed self-similar,
we collapse them on top of each other by plotting them versus $k\xi(t)$.
Here, $\xi(t)=\EEEE^{-1}\int k^{-1} E(k,t)\,\dd k$ is the weighted
integral of $k^{-1}$, where the spectra are normalized such that
$\int E(k,t)\,\dd k=\EEEE(t)$ is the magnetic energy density.
This ensures that the maxima of $E(k\xi)$ are always approximately near
$k\xi(t)=1$.
In addition, we must also compensate for the decay in amplitude
by multiplying the spectra by a time-dependent function, e.g.,
$\xi(t)^\beta$, where $\beta$ is a suitable exponent, so that the
compensated spectra all have the same height.
In this way, we find a universal spectral function by plotting
\begin{equation}
\left[\xi(t)\right]^\beta E\big(k\xi(t),t\big)\equiv\phi\big(k\xi(t)\big).
\label{Collapse}
\end{equation}
As an example, we show in \Fig{pkt_few_comp_Hf0_t2em5_k180a} the compensated
spectra for Run~B of B20, which we discuss in more detail below.
At this point, we just note that these were solutions to \Eq{Hall},
where the initial condition consists of a nonhelical magnetic field
with a spectrum $E(k)\propto k^4$ for $k\ll k_0$, with $k_0$ being the
initial peak wavenumber.
For $k\gg k_0$, we assume a decaying spectrum, here with $E(k)\propto
k^{-5/3}$, although this particular choice of the exponent was not important.
After some time, the spectral slopes of both subranges change:
At small $k$, the spectrum steepens from $k^4$ to $k^5$.
Beyond the peak, it falls off with a $k^{-7/3}$ inertial range, as was
already found by \cite{Bis96}.

To determine the theoretically expected value of $\beta$, we invoke the
condition that the compensated spectra be invariant under rescaling,
$t\to\tau t'$, $\xx\to\tau^q\xx'$, where $\tau$ is an arbitrary scaling
factor.
We recall that the dimensions of $E(k,t)$ are $[x]^5[t]^{-2}$, so
rescaling yields a factor $\tau^{5q-2}$.
In addition, expressing $E'$ in terms of its universal spectral function
$\phi(k\xi)$, the factor $\xi^\beta$ on the left-hand side of the
\Eq{Collapse} produces a factor $\xi^{-\beta}$ on the right-hand side
of \Eq{Collapse}, and therefore, after rescaling, a $\tau^{-\beta q}$
factor, i.e.,
\begin{equation}
E(k)\to\tau^{5q-2}\,E'(k)\propto
\xi(t)^{-\beta}\tau^{-\beta q}\phi(k\xi).
\end{equation}
Therefore, $5q-2=-\beta q$ must be satisfied in order that the compensated
spectra remain invariant under rescaling.
Thus, $\beta=2/q-5$, as already found in B20.
Inserting now $q=4/13$ yields $\beta=3/2$.
The total magnetic energy density is therefore
\begin{equation}
\EEEE=\int \xi^{-\beta} \phi(k\xi)\,\dd k=
\xi^{-(\beta+1)} \int \phi(k\xi)\,\dd(k\xi)
\propto t^{-(\beta+1)\,q},
\end{equation}
and since $\EEEE\propto t^{-p}$, we have\footnote{An equivalent, but
conceptually simpler route to the $t^{-10/13}$ decay law, suggested
by David Hosking (private communication), is to use \Eq{Hall} to argue
that the timescale $t_{\rm dec}$ for magnetic decay is proportional to
$\xi^2/B$.  Since $t_{\rm dec}$ is proportional to $t$ for selfsimilar
decay, we get, using $\xi^5 B^4=[B^2 (\xi^2/B)^{10/13}]^{13/4} =(B^2
t^{10/13})^{13/4}=\const$, the scaling $B^2\propto t^{-10/13}$.
The difference between MHD and Hall cascade is that, in the former,
$t_{\rm dec}$ is proportional to $\xi/B$, but $\propto\xi^2/B$ in
the latter.}
$p=(\beta+1)\,q$.
Using $\beta=2/q-5$, we have $p=2\,(1-2q)$; see Eq.~(28) of B20.
For $q=4/13\approx0.31$, we have $p=2\,(1-8/13)=10/13\approx0.78$,
which is not quite as close to the value reported in B20 as that of $q$,
but this could be ascribed to the lack of scale separation and also
the magnetic field no longer being strong enough so that the Lundquist
number,\footnote{Note that, unlike the case of MHD, in the present case
of Hall cascade, no wavenumber factor enters in the definition of the
Lundquist number.}
\begin{equation}
\Lu=\Brms/en_e\mu_0\eta,
\end{equation}
is no longer in the asymptotic regime.
This also resulted in the empirical value of $\beta$ being slightly larger
than the theoretical one, as we will see next.

\section{Comparison with simulations for different diffusivities}

In B20, various simulations of the Hall cascade have been presented,
including forced and decaying simulations, helical and nonhelical ones,
with constant and time-varying magnetic diffusivities, with and without
stratification, etc.
The main purpose of that work was to understand the dissipative losses
that would lead to resistive heating in the crust of a neutron star.
One of those simulations is particularly relevant for the present paper: his
Run~B, which had a relatively strong initial magnetic field, no helicity,
large scale-separation, and a magnetic diffusivity that decreased with
time in a power law fashion, allowing the simulation to retain a higher
Lundquist number as the magnetic field decreases.

In the present paper, we analyze his Run~B, which is here called Run~B1.
It is actually a new run, because we now have calculated the Hosking
integral during run time.
We also compare with another run (Run~B2), where we decreased the magnetic
diffusivity by a factor of two.
As in B20, $\eta$ is assumed to decrease with time proportional to $t^{-3/7}$.
We kept, however, the same resolution of $1024^3$ mesh points for both
runs, but we must keep in mind that this can lead to artifacts resulting
from a poorly resolved diffusive subrange for Run~B2.

\begin{table}
\begin{center}
\begin{tabular}{ccccccccccc}
Run& $\Lu$ & $t_1/[t]$ & $t_2/[t]$ & $\tilde{\eta}$ & $\tilde{B}_{\rm rms}$ & $\tilde{\epsilon}$ & $p$ & $q$ & $\beta$ & $p_{\rm H}$ \\
B1 &   650 &  0.2    &    500    &     0.024      &        600            &   $3\times10^6$    & $0.8\pm0.1$  & $0.3\pm0.1$  & $1.7\pm0.1$ & 0.16 \\
B2 &  1300 &   3     &    200    &     0.011      &        700            &   $6\times10^6$    & $0.78\pm0.05$& $0.31\pm0.05$& $1.6\pm0.05$& 0.11 \\
\end{tabular}
\caption{
Summary of runs discussed in this paper.
}\label{Tsummary}
\end{center}
\end{table}

In \Tab{Tsummary}, we compare several characteristic parameters: the
start and end times, $t_1$ and $t_2$, respectively, of the interval for
which averaged data have been accumulated, nondimensional measures of the
magnetic diffusivity, the magnetic field strength, and the dissipation,
$\tilde{\eta}$, $\tilde{B}_{\rm rms}$, and $\tilde{\epsilon}$,
respectively, and the instantaneous scaling exponents $p$, $q$, and
$\beta$.
For $\tilde{\eta}$, $\tilde{B}_{\rm rms}$, and $\tilde{\epsilon}$,
we compute the following time-averaged ratios:
\begin{equation}
\tilde{\eta}\equiv\bra{t\eta/\xi^2}_t,\quad
\tilde{B}_{\rm rms}\equiv\bra{\Brms/(en_e\mu_0\eta)}_t,\quad\mbox{and}\quad
\tilde{\epsilon}\equiv\bra{\epsilon/(e^2 n_e^2 \mu_0 \eta^3/\xi^2)}_t,
\end{equation}
where $\xi(t)={\cal E}^{-1}\int k^{-1} E(k,t)\,\dd k$ is the correlation
length and $\epsilon=\eta\mu_0\bra{\JJ^2}$ is the magnetic
dissipation with $\eta=\eta(t)$, as noted above.
These were also computed in B20.
Time is given in diffusive units, $[t]=(\eta k_0^2)^{-1}$.
In the runs of series B of B20, the value of $k_0$ is 180 times larger
than the lowest wavenumber $k_1\equiv2\pi/L$ of our cubic domain of
size $L^3$.

\begin{figure}
\centering
\includegraphics[width=0.9\columnwidth]{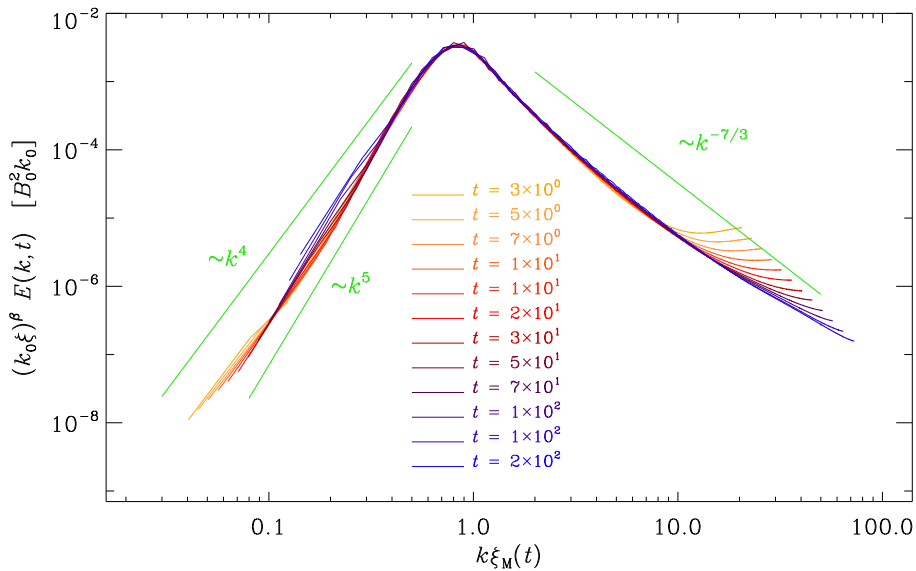}
\caption{
Compensated spectra for Run~B2.
}\label{pkt_few_comp_Hf0_t1em5_k180a_Hosk}
\end{figure}

\begin{figure}
\centering
\includegraphics[width=0.9\columnwidth]{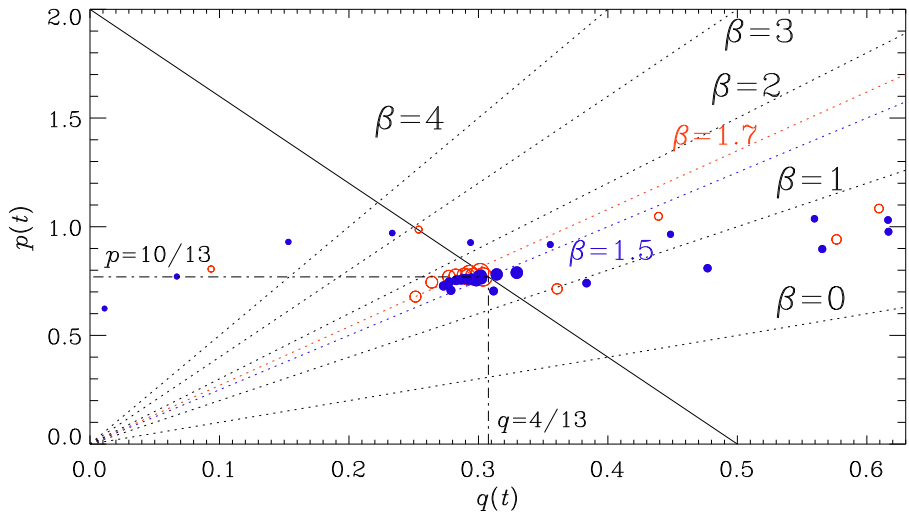}
\caption{
$pq$ diagrams for Runs~B1 (open red symbols) and B2 (closed blue symbols).
Larger symbols denote later times.
}\label{pq_comp}
\end{figure}

\begin{figure}
\centering
\includegraphics[width=0.9\columnwidth]{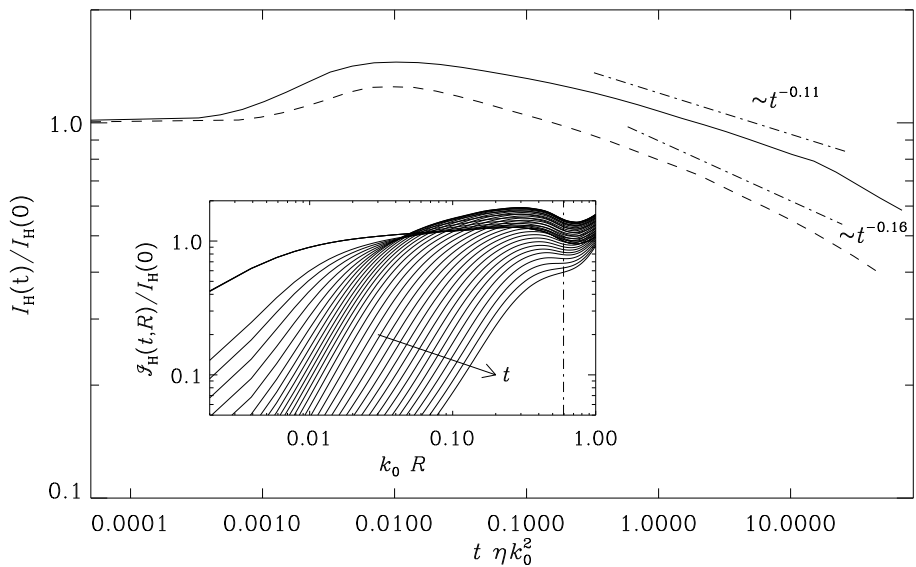}
\caption{
Evolution of $I_{\rm H}(t)$, showing first a slight increase and then
a decline proportional to $t^{-0.16}$ for Run~B1 and proportional to
$t^{-0.11}$ for Run~B2.
The inset shows the evolution of ${\cal I}_{\rm H}(R;t)$ as a function
of $R$ for increasing values of $t$ (indicated by the arrow) for Run~B2.
The abscissae of the main plot and the inset are normalized by
$\eta k_0^2$ and $k_0$, respectively.
}\label{psaff_comp}
\end{figure}

It turns out that a lower resistivity is important for obtaining
the expected scaling.
We therefore now consider Run~B2, where $\Lu\approx1300$.
The result is shown in \Fig{pkt_few_comp_Hf0_t1em5_k180a_Hosk}, where
we used $\beta=1.6$ as the best fit, which is still slightly larger than
the expected value of 3/2, but it goes in the right direction.
Therefore, we show in \Fig{pkt_few_comp_Hf0_t1em5_k180a_Hosk} the
resulting compensated spectra for Run~B2, where the magnetic diffusivity
is half that of Run~B1 and $\Lu$ is now twice as large a before;
see \Tab{Tsummary}.
We use the {\sc Pencil Code} \citep{PC21} and, in both cases, we use a
resolution of $1024^3$ mesh points.

Another comparison between Runs~B1 and B2 is shown in \Fig{pq_comp},
where we compare their evolution in the $pq$--diagram.
While Run~B1 clearly evolves along the $\beta\approx1.7$ line,
Run~B2 tends to be closer to the $\beta\approx3/2$ line.
Note also that both runs settle near the $p=2(1-2q)$ self-similarity
line (B20), although we begin to see departures near the end of the run,
which is due to the finite size of the domain.

Finally, we show in \Fig{psaff_comp} the scalings of $I_{\rm H}(t)$,
where we see that the decay exponent $p_{\rm H}\equiv-\dd\ln I_{\rm H}/\dd\ln t$
is about $p_{\rm H}\approx0.16$ for Run~B1 and about $0.11$ for Run~B2.
Earlier work by \cite{ZSB22} showed that $p_{\rm H}$ decreases as
the Lundquist number increases, and is, in MHD, around 0.2 for
$\Lu\approx10^3$, and decreases to $p_{\rm H}\approx0.01$ for
$\Lu\approx4\times10^7$.
Such large values can currently only be obtained with magnetic
hyper-diffusivity \citep{HS21, ZSB22}, but this has not been attempted
in the present work.

\begin{table}
\begin{center}
\begin{tabular}{lllll}
MHD  & $[B]=[x]/[t]$   & $[\bra{h}]=[x]^3/[t]^2$ & $q=2/3$ \; (helical) \\[.4mm]
     &                 & $[I_{\rm H}]=[x]^9/[t]^4$ & $q=4/9$ \; (nonhel) \\[.4mm]
     & $[E(k,t)]=[x^3][t]^{-2}$ & $\beta=2/q-3$  & $\beta=2\cdot9/4-3=3/2$ \\[.4mm]
     &                          & $p=2\,(1-q)$  & $p=2\,(1-4/9)=10/9$ \\[2mm]
Hall & $[B]=[x]^2/[t]$ & $[\bra{h}]=[x]^5/[t]^2$ & $q=2/5$ \; (helical) \\[.4mm]
     &                 & $[I_{\rm H}]=[x]^{13}/[t]^4$ & $q=4/13$ \; (nonhel) \\[.4mm]
     & $[E(k,t)]=[x^5][t]^{-2}$ & $\beta=2/q-5$  & $\beta=2\cdot13/4-5=3/2$ \\[.4mm]
     &                          & $p=2\,(1-2q)$  & $p=2\,(1-2\cdot4/13)=10/13$\\[1mm]
\end{tabular}
\caption{
Comparison of the scalings for MHD and the Hall cascade.
}\label{Tcomparison}
\end{center}
\end{table}

As already noted by \cite{ZSB22}, there is an initial increase in
$I_{\rm H}(t)$.
This is explained by the fact that the magnetic field obeys Gaussian
statistics initially, but not during the later evolution.
The inset shows the $R$ dependence of ${\cal I}_{\rm H}(R;t)$ for
different $t$.
The relevant value of $R$ is deemed to be at the location where the
local slope of ${\cal I}_{\rm H}(R)$ is minimum at late times.

\section{Conclusion}
\label{Concl}

The present work has highlighted the power of dimensional
arguments, which were here applied to the case of the Hall cascade without
helicity, where the magnetic field is naturally represented as a quantity
with units of a magnetic diffusivity.
The magnetic helicity density has units of $\m^5\s^{-2}$ and the
Hosking integral has units of $\m^{13}\s^{-4}$, which yields $q=4/13$,
$\beta=3/2$, and $p=10/13$.
Comparing with standard MHD, where the magnetic field has units of
$\m\s^{-1}$, our exponents $p$ and $q$ are now smaller, but $\beta$
is still the same in both cases; see \Tab{Tcomparison} for a comparison
between Hall cascade and MHD.
The empirically determined value of $\beta$ is somewhat larger, but this
can be explained by finite scale separation and small Lundquist numbers.

The decay properties of the Hall cascade are important in understanding
resistive heating in neutron stars while producing at the same time larger
scale magnetic fields at a certain speed through inverse cascading (B20).
Such simulations have already been done in spherical geometry 
\citep{Gourgouliatos+20}, but the magnetic field in those simulations
did not yet exhibit clear forward or inverse cascading.
This is presumably due to their initial magnetic field being strongly
localized at intermediate length scales.
Using an initial broken power law, as done here, would help producing
the expected forward or inverse cascadings, but this may also require
much larger resolution than what is currently possible.
Similarly, of course, the values of $n_e$ and $\eta$ are depth dependent
in real neutron stars, but the work of B20 showed that this did not
affect the scaling behavior of the magnetic decay.
Therefore, the importance of the Hosking integral may well carry over
to real neutron stars.

The possible role of reconnection in the Hall cascade remains still an
open question.
In the case of MHD, reconnection has been discussed extensively by
\cite{HS21}, making reference to earlier work by \cite{Zhou+19, Zhou+20}
and \cite{Bhat+21}.
Also in the Hall cascade there is the possibility that the decay of
$\EEEE$ could be slowed down in an intermediate range of values of the
Lundquist number.
As shown in \cite{ZSB22}, this could lead to a termination line in the
$pq$ diagrams that is different from the $p=2(1-2q)$ self-similarity
line discussed here.
At the moment, however, there is no compulsory evidence for deviations
of the solutions from the self-similarity line.

\section*{Acknowledgements}
I would like to thank the Isaac Newton Institute for Mathematical
Sciences, Cambridge, for support and hospitality during the programme
``Frontiers in dynamo theory: from the Earth to the stars'' where the
work on this paper was undertaken.
I am also grateful to the two referees for their comments, which have
led to improvements in the presentation.

\section*{Funding}
This work was supported by EPSRC grant no EP/R014604/1 and
the Swedish Research Council (Vetenskapsr{\aa}det, 2019-04234).
Nordita is sponsored by Nordforsk.
We acknowledge the allocation of computing resources provided by the
Swedish National Allocations Committee at the Center for Parallel
Computers at the Royal Institute of Technology in Stockholm and
Link\"oping.

\section*{Declaration of Interests}
The authors report no conflict of interest.

\section*{Data availability statement}
The data that support the findings of this study are openly available
on Zenodo at doi:10.5281/zenodo.7357799 (v2022.11.24).
All calculations have been performed with the {\sc Pencil Code}
\citep{PC21}; DOI:10.5281/zenodo.3961647.

\section*{Author ORCID}
A. Brandenburg, https://orcid.org/0000-0002-7304-021X

\bibliographystyle{jpp}
\bibliography{ref}

\begin{thebibliography}{19}
\expandafter\ifx\csname natexlab\endcsname\relax\def\natexlab#1{#1}\fi
\def\au#1{#1} \def\ed#1{#1} \def\yr#1{#1}\def\at#1{#1}\def\jt#1{\textit{#1}}
  \def\bt#1{#1}\def\bvol#1{\textbf{#1}} \def\vol#1{#1} \def\pg#1{#1}
  \def\publ#1{#1}\def\arxiv#1{#1}\def\org#1{#1}\def\st#1{\textit{#1}}

\bibitem[{Anzuini} {\em et~al.\/}(2022){Anzuini}, {Melatos}, {Dehman},
  {Vigan{\`o}} \& {Pons}]{Anzuini+22}
{\sc \au{{Anzuini}, F.}, \au{{Melatos}, A.}, \au{{Dehman}, C.},
  \au{{Vigan{\`o}}, D.} \& \au{{Pons}, J.~A.}} \yr{2022}  \at{{Thermal
  luminosity degeneracy of magnetized neutron stars with and without hyperon
  cores}}.  \jt{\mnras}  \bvol{515}~(2),  \pg{3014--3027},  \arxiv{arXiv:
  2205.14793}.

\bibitem[{Bhat} {\em et~al.\/}(2021){Bhat}, {Zhou} \& {Loureiro}]{Bhat+21}
{\sc \au{{Bhat}, Pallavi}, \au{{Zhou}, Muni} \& \au{{Loureiro}, Nuno~F.}}
  \yr{2021}  \at{{Inverse energy transfer in decaying, three-dimensional,
  non-helical magnetic turbulence due to magnetic reconnection}}.  \jt{\mnras}
  \bvol{501}~(2),  \pg{3074--3087},  \arxiv{arXiv: 2007.07325}.

\bibitem[{Biskamp} {\em et~al.\/}(1996){Biskamp}, {Schwarz} \& {Drake}]{Bis96}
{\sc \au{{Biskamp}, D.}, \au{{Schwarz}, E.} \& \au{{Drake}, J.~F.}} \yr{1996}
  \at{{Two-Dimensional Electron Magnetohydrodynamic Turbulence}}.  \jt{\prl}
  \bvol{76}~(8),  \pg{1264--1267}.

\bibitem[{Brandenburg}(2020)]{B20}
{\sc \au{{Brandenburg}, Axel}} \yr{2020}  \at{{Hall Cascade with Fractional
  Magnetic Helicity in Neutron Star Crusts}}.  \jt{\apj}  \bvol{901}~(1),
  \pg{18},  \arxiv{arXiv: 2006.12984}.

\bibitem[{Brandenburg} \& {Kahniashvili}(2017)]{BK17}
{\sc \au{{Brandenburg}, Axel} \& \au{{Kahniashvili}, Tina}} \yr{2017}
  \at{{Classes of Hydrodynamic and Magnetohydrodynamic Turbulent Decay}}.
  \jt{\prl}  \bvol{118}~(5),  \pg{055102},  \arxiv{arXiv: 1607.01360}.

\bibitem[{Brandenburg} {\em et~al.\/}(2015){Brandenburg}, {Kahniashvili} \&
  {Tevzadze}]{BKT15}
{\sc \au{{Brandenburg}, Axel}, \au{{Kahniashvili}, Tina} \& \au{{Tevzadze},
  Alexander~G.}} \yr{2015}  \at{{Nonhelical Inverse Transfer of a Decaying
  Turbulent Magnetic Field}}.  \jt{\prl}  \bvol{114}~(7),  \pg{075001},
  \arxiv{arXiv: 1404.2238}.

\bibitem[{Cho} \& {Lazarian}(2009)]{CL09}
{\sc \au{{Cho}, Jungyeon} \& \au{{Lazarian}, A.}} \yr{2009}  \at{{Simulations
  of Electron Magnetohydrodynamic Turbulence}}.  \jt{\apj}  \bvol{701}~(1),
  \pg{236--252},  \arxiv{arXiv: 0904.0661}.

\bibitem[{Goldreich} \& {Reisenegger}(1992)]{GR92}
{\sc \au{{Goldreich}, Peter} \& \au{{Reisenegger}, Andreas}} \yr{1992}
  \at{{Magnetic Field Decay in Isolated Neutron Stars}}.  \jt{\apj}
  \bvol{395},  \pg{250}.

\bibitem[{Gourgouliatos} \& {Hollerbach}(2018)]{Gourgouliatos+18}
{\sc \au{{Gourgouliatos}, Konstantinos~N.} \& \au{{Hollerbach}, Rainer}}
  \yr{2018}  \at{{Magnetic Axis Drift and Magnetic Spot Formation in Neutron
  Stars with Toroidal Fields}}.  \jt{\apj}  \bvol{852}~(1),  \pg{21},
  \arxiv{arXiv: 1710.01338}.

\bibitem[{Gourgouliatos} {\em et~al.\/}(2018){Gourgouliatos}, {Hollerbach} \&
  {Archibald}]{Gourgouliatos+18b}
{\sc \au{{Gourgouliatos}, Konstantinos~N.}, \au{{Hollerbach}, Rainer} \&
  \au{{Archibald}, Robert~F.}} \yr{2018}  \at{{Modelling neutron star magnetic
  fields}}.  \jt{Astron. Geophys.}  \bvol{59}~(5),  \pg{5.37--5.42}.

\bibitem[{Gourgouliatos} {\em et~al.\/}(2020){Gourgouliatos}, {Hollerbach} \&
  {Igoshev}]{Gourgouliatos+20}
{\sc \au{{Gourgouliatos}, Konstantinos~N.}, \au{{Hollerbach}, Rainer} \&
  \au{{Igoshev}, Andrei~P.}} \yr{2020}  \at{{Powering central compact objects
  with a tangled crustal magnetic field}}.  \jt{\mnras}  \bvol{495}~(2),
  \pg{1692--1699},  \arxiv{arXiv: 2005.02410}.

\bibitem[{Gourgouliatos} {\em et~al.\/}(2016){Gourgouliatos}, {Wood} \&
  {Hollerbach}]{Gourgouliatos+16}
{\sc \au{{Gourgouliatos}, Konstantinos~N.}, \au{{Wood}, Toby~S.} \&
  \au{{Hollerbach}, Rainer}} \yr{2016}  \at{{Magnetic field evolution in
  magnetar crusts through three-dimensional simulations}}.  \jt{Proc. Nat.
  Acad. Sci.}  \bvol{113}~(15),  \pg{3944--3949},  \arxiv{arXiv: 1604.01399}.

\bibitem[{Hosking} \& {Schekochihin}(2021)]{HS21}
{\sc \au{{Hosking}, David~N.} \& \au{{Schekochihin}, Alexander~A.}} \yr{2021}
  \at{{Reconnection-Controlled Decay of Magnetohydrodynamic Turbulence and the
  Role of Invariants}}.  \jt{Phys. Rev. X}  \bvol{11}~(4),  \pg{041005},
  \arxiv{arXiv: 2012.01393}.

\bibitem[{Hosking} \& {Schekochihin}(2022)]{HS22}
{\sc \au{{Hosking}, David~N.} \& \au{{Schekochihin}, Alexander~A.}} \yr{2022}
  \at{{Cosmic-void observations reconciled with primordial magnetogenesis}} ,
  \arxiv{arXiv: 2203.03573}.

\bibitem[{Igoshev} {\em et~al.\/}(2021){Igoshev}, {Popov} \&
  {Hollerbach}]{Igoshev+21}
{\sc \au{{Igoshev}, Andrei~P.}, \au{{Popov}, Sergei~B.} \& \au{{Hollerbach},
  Rainer}} \yr{2021}  \at{{Evolution of Neutron Star Magnetic Fields}}.
  \jt{Universe}  \bvol{7}~(9),  \pg{351},  \arxiv{arXiv: 2109.05584}.

\bibitem[{Pencil Code Collaboration} {\em et~al.\/}(2021){Pencil Code
  Collaboration}, {Brandenburg}, {Johansen}, {Bourdin}, {Dobler}, {Lyra},
  {Rheinhardt}, {Bingert}, {Haugen}, {Mee}, {Gent}, {Babkovskaia}, {Yang},
  {Heinemann}, {Dintrans}, {Mitra}, {Candelaresi}, {Warnecke},
  {K{\"a}pyl{\"a}}, {Schreiber}, {Chatterjee}, {K{\"a}pyl{\"a}}, {Li},
  {Kr{\"u}ger}, {Aarnes}, {Sarson}, {Oishi}, {Schober}, {Plasson}, {Sandin},
  {Karchniwy}, {Rodrigues}, {Hubbard}, {Guerrero}, {Snodin}, {Losada},
  {Pekkil{\"a}} \& {Qian}]{PC21}
{\sc \au{{Pencil Code Collaboration}}, \au{{Brandenburg}, Axel},
  \au{{Johansen}, Anders}, \au{{Bourdin}, Philippe}, \au{{Dobler}, Wolfgang},
  \au{{Lyra}, Wladimir}, \au{{Rheinhardt}, Matthias}, \au{{Bingert}, Sven},
  \au{{Haugen}, Nils}, \au{{Mee}, Antony}, \au{{Gent}, Frederick},
  \au{{Babkovskaia}, Natalia}, \au{{Yang}, Chao-Chin}, \au{{Heinemann},
  Tobias}, \au{{Dintrans}, Boris}, \au{{Mitra}, Dhrubaditya},
  \au{{Candelaresi}, Simon}, \au{{Warnecke}, J{\"o}rn}, \au{{K{\"a}pyl{\"a}},
  Petri}, \au{{Schreiber}, Andreas}, \au{{Chatterjee}, Piyali},
  \au{{K{\"a}pyl{\"a}}, Maarit}, \au{{Li}, Xiang-Yu}, \au{{Kr{\"u}ger}, Jonas},
  \au{{Aarnes}, J{\o}rgen}, \au{{Sarson}, Graeme}, \au{{Oishi}, Jeffrey},
  \au{{Schober}, Jennifer}, \au{{Plasson}, Rapha{\"e}l}, \au{{Sandin},
  Christer}, \au{{Karchniwy}, Ewa}, \au{{Rodrigues}, Luiz}, \au{{Hubbard},
  Alexander}, \au{{Guerrero}, Gustavo}, \au{{Snodin}, Andrew}, \au{{Losada},
  Illa}, \au{{Pekkil{\"a}}, Johannes} \& \au{{Qian}, Chengeng}} \yr{2021}
  \at{{The Pencil Code, a modular MPI code for partial differential equations
  and particles: multipurpose and multiuser-maintained}}.  \jt{J. Open Source
  Softw.}  \bvol{6}~(58),  \pg{2807},  \arxiv{arXiv: 2009.08231}.

\bibitem[{Zhou} {\em et~al.\/}(2022){Zhou}, {Sharma} \& {Brandenburg}]{ZSB22}
{\sc \au{{Zhou}, Hongzhe}, \au{{Sharma}, Ramkishor} \& \au{{Brandenburg},
  Axel}} \yr{2022}  \at{{Scaling of the Hosking integral in decaying
  magnetically dominated turbulence}}.  \jt{J. Plasma Phys.}  \bvol{88}~(6),
  \pg{905880602},  \arxiv{arXiv: 2206.07513}.

\bibitem[{Zhou} {\em et~al.\/}(2019){Zhou}, {Bhat}, {Loureiro} \&
  {Uzdensky}]{Zhou+19}
{\sc \au{{Zhou}, Muni}, \au{{Bhat}, Pallavi}, \au{{Loureiro}, Nuno~F.} \&
  \au{{Uzdensky}, Dmitri~A.}} \yr{2019}  \at{{Magnetic island merger as a
  mechanism for inverse magnetic energy transfer}}.  \jt{Phys. Rev. Res.}
  \bvol{1}~(1),  \pg{012004},  \arxiv{arXiv: 1901.02448}.

\bibitem[{Zhou} {\em et~al.\/}(2020){Zhou}, {Loureiro} \& {Uzdensky}]{Zhou+20}
{\sc \au{{Zhou}, Muni}, \au{{Loureiro}, Nuno~F.} \& \au{{Uzdensky}, Dmitri~A.}}
  \yr{2020}  \at{{Multi-scale dynamics of magnetic flux tubes and inverse
  magnetic energy transfer}}.  \jt{J. Plasma Phys.}  \bvol{86}~(4),
  \pg{535860401},  \arxiv{arXiv: 2001.07291}.

\end{thebibliography}
\end{document}